\documentclass{aa}

\usepackage{graphicx}
\usepackage{txfonts}
\usepackage[separate-uncertainty=true]{siunitx} 

\usepackage{natbib}
\bibpunct{(}{)}{;}{a}{}{,}
\usepackage{hyperref}
\hypersetup{colorlinks=true, allcolors=blue}

\newcommand{\cov}[2]{\langle \vec{#1} \vec{{#2}^\mathrm{T}} \rangle} 
\newcommand{\covnv}[2]{\langle {#1} {{#2}} \rangle}
\newcommand{\var}[2]{\langle \vec{#1^2_{\mathrm{#2}}} \rangle} 
\newcommand{\varnv}[2]{\langle {#1^2_{#2}} \rangle} 
\newcommand{\mat}[1]{\mathit{#1}}

\begin{document}


   \title{Integrated turbulence parameters' estimation from NAOMI adaptive optics telemetry data \thanks{Based on observations collected at the European Southern Observatory under ESO programme 60.A-9278(D).}}

   \author{Nuno Morujão \inst{1,2}
          \and
          Carlos Correia \inst{1,2,3}
          \and
          Paulo Andrade \inst{1,2}
          \and
          Julien Woillez\inst{4}
          \and
          Paulo Garcia \inst{1,2}
          }

   \institute{Faculdade de Engenharia da Universidade do Porto, Rua Dr. Roberto Frias, s/n, 4200-465 Porto, Portugal\\
            \and
            CENTRA – Centro de Astrofísica e Gravitação, IST, Universidade de Lisboa, 1049-001 Lisboa, Portugal \\
            \and
            Space ODT – Optical Deblurring Technologies Unip Lda, 4050-277 Porto, Portugal\\
            \and
            European Southern Observatory, Garching bei Muenchen, Germany
             }

   \date{}

  \abstract 
   {Monitoring turbulence parameters is crucial in high-angular resolution astronomy for various purposes, such as optimising adaptive optics systems or fringe trackers. The former systems are present at most modern observatories and will remain significant in the future. This makes them a valuable complementary tool for the estimation of turbulence parameters. } 
   {The feasibility of estimating turbulence parameters from low-resolution sensors remains untested. We performed seeing estimates for both simulated and on-sky telemetry data sourced from the new adaptive optics module installed on the four Auxiliary Telescopes of the Very Large Telescope Interferometer.}
   {The seeing estimates were obtained from a modified and optimised algorithm that employs a chi-squared modal fitting approach to the theoretical von Kármán model variances. The algorithm was built to retrieve turbulence parameters while simultaneously estimating and accounting for the remaining and measurement error. A Monte Carlo method was proposed for the estimation of the statistical uncertainty of the algorithm.}
   {The algorithm is shown to be able to achieve per-cent accuracy in the estimation of the seeing with a temporal horizon of $\SI{20}{\second}$ on simulated data. A ${(\SI{0.76}{\arcsecond} \pm 1.2\%|_\mathrm{stat} \pm 1.2\%|_\mathrm{sys})}$ median seeing was estimated from on-sky data collected from 2018 to 2020. The spatial distribution of the Auxiliary Telescopes across the Paranal Observatory was found to not play a role in the value of the seeing.}
   {}
   
   \titlerunning{Turbulence parameters estimation from NAOMI AO telemetry data}
   \authorrunning{Nuno Morujão et al.}
   \maketitle
   
%

\section{Introduction} 
\label{sec: intro}

Earth's atmosphere is widely understood as a turbulent fluid with local variations in temperature, density, and velocity that translate into fluctuations in the refractive index of the air. The changes in the refractive index impact the propagation of electromagnetic radiation \citep{roggemann1996imaging, wyngaard2010turbulence}. In strong turbulence, the electric field's amplitude and phase are affected, causing scintillation and blur. When the turbulence affects only the phase, we are in a low-turbulence environment; this is the case for this work. 

The von Kármán model \citep{Karman1948} is widely used in adaptive optics (AO) as a descriptive model of the energy spectrum of turbulence and it describes the fluctuation of the refractive index \citep{Avila2021,Doelman2020,Fetick2018}. This model uses two turbulence parameters, the Fried parameter, $r_0$, and the outer scale, $\mathcal{L}_0$, to describe the average variance of the phase difference between two points separated by a distance, $\rho$, through a phase structure function, $D_{\vec{\vec{\phi}}}$. The model is used to ascertain the effects of the atmosphere on angular resolution through the seeing \citep{hardy1998adaptive}. The $r_0$ indicates the separation over which a wavefront is aberrated by roughly a one-radian root mean square \citep{Fried1965}. The outer scale dictates the inertial region of turbulence \citep{Voitsekhovich1995}. It serves as a saturation parameter, which is absent in the Kolmogorov structure function \citep{Kolmogorov1941}, and bounds the energy of the whole model.

The tracking of turbulence parameters is of importance to high-resolution astronomy:

\begin{itemize}
    \item in system commissioning and optimisation \citep{ESPRESSO2021};
    \item in the reconstruction of the point-spread function (PSF) \citep{WAGNER2022,Martin2019};
    \item in the optimisation of fringe trackers \citep{GRAVITY2019};
    \item in site testing and characterisation \citep{Zhu2023,Tillayev2021,Liu2015,Rami2012};
    \item in the tracking of climate-induced changes \citep{Kooten2022,cantalloube2020}; 
    \item in AO pre-compensation for free-space optical communications \citep{Griffiths2023,Osborn2021}.
\end{itemize}

Turbulence parameter tracking campaigns at Paranal, traditionally done with a MASS-DIMM system, have shown a need for complementary seeing estimates \citep{Butterley2020,Osborn2018,Masciadri2014,sarazin2008seeing} due to non-consensual estimations between instruments. Complementary systems to MASS-DIMM are currently an open research question, and various system architectures have been put forward \citep{Perera2023,Tokovinin2021,Guesalaga2021}.

The use of AO telemetry data for the estimation of integrated turbulence parameters is one such alternative \citep{Jolissaint2018,Fusco2004}. Adaptive-optics systems are ubiquitous in current and future-generation telescopes. Estimating turbulence parameters from telemetry data has the advantage of sourcing data from already-built systems. Additionally, since the processed AO telemetry is sourced directly from the observing telescopes, we have a line-of-sight seeing estimation for each telescope. Finally, since the telemetry encompasses the full light path from the guide star to the telescope, we can include the dome turbulence contribution to the seeing. The former is particularly interesting to high-resolution imaging since it has been shown to greatly impact image quality \citep{Munro2023,Lai2019,Salmon2009}.

Testing the algorithm on simulated and on-sky data is crucial to validate the estimation method. This study utilises the telemetry data produced by the new AO module of the Very Large Telescope Interferometer (VLTI), NAOMI \citep{Woillez2019}, to achieve this validation. NAOMI is installed on all four Auxiliary Telescopes (ATs) of the VLTI and can provide up to four concurrent telemetry samples. However, as a low-resolution system with a $4\times4$ Shack-Hartmann wavefront sensor (SH-WFS), it is a challenging scenario for estimating turbulence parameters from AO telemetry since only low-order modal variances are available for the fit. These additional complications affect the estimation of the turbulence parameters, and their impact is not yet understood.

We aim to conduct a quantitative validation and optimisation of the estimation algorithm under typical observing conditions at the Paranal Observatory by applying it to simulated data. Additionally, the analysis of real telemetry data (discussed in Section \ref{sec: results}) aims to retrieve turbulence parameters observed at the Paranal Observatory from 2018 to 2020. This information will help to assess:
\begin{itemize}
    \item the performance of the algorithm in the estimation of the turbulence parameters from on-sky telemetry;
    \item the distribution of the seeing across the Paranal Observatory;
    \item the algorithm performance in comparison with currently employed seeing trackers.
\end{itemize}

In Section \ref{sec: methods}, the modal wavefront representation and reconstruction are detailed. The reconstruction challenges are described. The theoretical models for retrieving pseudo-open-loop modes and the fitting algorithm for estimating integrated turbulence parameters are presented. In Section \ref{subsec: simulation results}, a simulated pipeline of the NAOMI system is used to validate and optimise the algorithm. Section \ref{sec: results} applies the optimised estimation algorithm to $N=8170$ on-sky telemetry samples sourced from NAOMI. Conclusions are drawn in Section \ref{sec: conclusion}.


\section{Methods} 
\label{sec: methods}

\subsection{Reconstruction of Zernike coefficients}

We chose to represent the wavefront, $\vec{\vec{\phi}}$, as
\begin{equation}
    \vec{\vec{\phi}} = \sum^{M}_{i=2} a_i Z_i(\vec{\rho}),
\end{equation}

\noindent where $M$ is the total number of modes, $a_i$\footnote{If not explicitly stated in this article, a modal coefficient is implicitly represented at a fixed time, $t$.} are the modal coefficients, $Z_i(\vec{\rho})$ is the Zernike polynomial of Noll order $i$, and $\vec{\rho} = (\rho/R,\theta)$ represents the position vector in polar coordinates  \citep{Noll1976}. The piston contribution, $i=1$, has been discarded in the representation since it plays no role in imaging through a single pupil.

The NAOMI system tracked wavefront aberrations using a Shack-Hartmann sensor. Conventionally \citep{Southwell1980}, a set of slope measurements, $\vec{s}$, is related to the modal coefficients of the Zernike wavefront expansion through 

\begin{equation}
    \label{eq: zernike coeff rel SH}
    \vec{s} = [ s_1, \dots, s_{2N} ]^\mathrm{T} = \mat{G}\vec{a} + \vec{e},
\end{equation}

\noindent where a $N$ lenslet sensor is assumed. The gradient matrix, $\mat{G}$, translates the Zernike coefficients into Shack-Hartmann slopes, and $\vec{e}$ is the measurement error, following the same convention and notation as \cite{Dai1996}.

For modelling purposes, we truncated the phase representation to a finite set of Zernike coefficients and split it into two subsets of the reconstructed, $\vec{a_\mathrm{f}}$, and remaining modes, $\vec{a_\mathrm{r}}$. Equation (\ref{eq: zernike coeff rel SH}) can be rewritten as 

\begin{equation}
    \label{eq: truncated a eq}
    \vec{s} = \mat{G} \begin{bmatrix} \vec{a_\mathrm{f}} \\ \vec{a_\mathrm{r}} \end{bmatrix} + \vec{e} = \mat{H}\vec{a_\mathrm{f}} + \mat{G_\mathrm{r}}\vec{a_\mathrm{r}} + \vec{e},
\end{equation}

\noindent where $\mat{H}$ are the first $L$ columns of the $\mat{G}$ matrix, and $\mat{G_\mathrm{r}}$ was constructed from the remaining higher-order columns. The least-squares solution of equation (\ref{eq: truncated a eq}) is given as

\begin{equation}
    \label{eq: ls solution of b}
    \vec{b} = \mat{H^+}\vec{s},
\end{equation}

\noindent where $\vec{b} = [b_2, \dots, b_{L}]^\mathrm{T}$ are the reconstructed modal coefficients, and $\mat{H^+}$ is the reconstruction matrix obtained from the general inverse matrix of $\mat{H}$,

\begin{equation}
    \mat{H^+} = (\mat{H}^\mathrm{T}\mat{H})^{-1}\mat{H}^\mathrm{T}.
\end{equation}

\subsection{Pseudo-open slopes}

The former discussion is valid for an AO system operating in open loop. Instead, the AO systems we are interested in operate in a negative feedback closed loop \citep{hardy1998adaptive}. To retrieve the complete phase coefficients, $\vec{b}$, the control loop must be undone. Undoing the control loop corresponds to the summation of deformable mirror voltages, $\vec{v}_{\mathrm{DM}}$, with the corresponding Shack-Hartmann slope measurements, $\vec{s}_{\mathrm{SH}}$, in the modal space.

The modal coefficients for the deformable mirror, $\vec{c}_\mathrm{DM}$, were obtained from the mirror voltages and the matrix that translated the response of the deformable mirror into mode coefficients, $\mat{DM2M}$, through 

\begin{equation}
    \label{eq: rec commands}
    \vec{c}_{\mathrm{DM}}(t) = \mat{DM2M}\,\vec{v}_{\mathrm{DM}}(t).
\end{equation}

The residual modes, $\vec{b}_\mathrm{SH}$, were obtained from the residual slopes, $\vec{s}_{\mathrm{SH}}$, through the control matrix, $\mat{CM}$, that translated Shack-Hartmann slopes into Zernike modes using

\begin{equation}
    \label{eq: res slopes}
    \vec{b}_{\mathrm{SH}}(t-\tau) = \mat{CM}\,\vec{s}_{\mathrm{SH}}(t-\tau).
\end{equation}

\noindent A delay, $\tau$, was introduced in the slope reconstruction to account for the system's response time. In the case of NAOMI, the total loop delay is \SI{4.6}{\ms} \citep{Woillez2019}. The pseudo-open-loop coefficients were then 

\begin{equation}
    \vec{b}(t) = \vec{c}_\mathrm{DM}(t) + \vec{b}_\mathrm{SH}(t-\tau).
\end{equation}

\subsection{Finite representation limitations}

The relation between the true coefficients, $\vec{a}$, and reconstructed coefficients, $\vec{b}$, was obtained by substituting equation (\ref{eq: truncated a eq}) into equation (\ref{eq: ls solution of b}), such that

\begin{equation}
    \label{eq: finite rec of modes}
    \vec{b} = \vec{a_\mathrm{f}} + \mat{C}\,\vec{a_\mathrm{r}} + \mat{H^+}\,\vec{e},
\end{equation}

\noindent where $\mat{C}$ is the cross-talk matrix defined by \cite{Dai1996} as

\begin{equation}
    \mat{C} = \mat{H^+}\,\mat{G_\mathrm{r}}.
\end{equation}

The $\mat{C}$ matrix introduced a dependence on the remaining modal coefficients in two ways: i) a geometric coupling due to the non-orthogonality of the Zernike polynomial derivatives \citep{Herrmann1981} and ii) the aliasing term. Although these are two different effects, they were treated as one -- the remaining error, $\vec{R}$. 

\subsection{Zernike coefficient variances}

Estimating the Zernike modal variances requires the computation of the remaining error, which requires knowledge of the remaining modal variances. These were estimated through analytical expressions. The theoretical values for the von Kármán (co)-variance, $\cov{a_\mathrm{r}}{a_\mathrm{r}}$\footnote{ $\langle .\rangle$ indicates the ensemble average of a parameter.}, were obtained through the expressions derived by \cite{takato1995spatial} and they are themselves a function of the Fried parameter, $r_0$, and the outer scale, $\mathcal{L}_0$.

The reconstruction also assumes that the measurement errors are equal and uncorrelated between measurements \citep{Southwell1980}. As a consequence the variance of the measurement error, $\var{\vec{e}}{}$, can be treated as a constant variance across slopes, such that

\begin{equation}
    \label{eq: noise def}
    \varnv{{e}}{i} = {\sigma_0^2}.
\end{equation}

The Zernike coefficient variances were evaluated from equation (\ref{eq: finite rec of modes}) through

\begin{equation}
    \var{\vec{b}}{} = \var{\vec{a_\mathrm{f}}}{} + \mat{C}\,\cov{a_\mathrm{r}}{a_\mathrm{r}}\,\mat{C}^\mathrm{T} + 2\mat{C}\cov{a_\mathrm{r}}{a_\mathrm{f}} + \mat{H^+}\,\var{\vec{e}}{}\,(\mat{H^+})^\mathrm{T},
\end{equation}

\noindent where $\var{\vec{b}}{}$ represents the variance of the reconstructed modal coefficients and $\cov{\vec{a_i}}{\vec{a_j}}$ the (co-)variances of the exact modal coefficients. The measurement error is assumed to be uncorrelated from the modal coefficients (i.e. $\cov{e}{a_\mathrm{r}} = \cov{e}{a_\mathrm{f}} = 0$). The variance of the reconstructed modes is then
\begin{equation}
    \label{eq: modal variances}
    \var{\vec{b}}{} = \var{\vec{a}}{f} + \vec{R} + \vec{\sigma^2_{\mathrm{n}}},
\end{equation}

\noindent where $\vec{\sigma^2_{\mathrm{n}}}$ is associated with the measurement error in the Zernike space and is expressed as

\begin{equation}
    \label{eq: measurement error}
    {{\sigma^2_{\mathrm{n},i}}}= \sigma_0^2 \left[\mat{H}^\mathrm{T}\mat{H}\right]_{i, i}^{-1},
\end{equation}

\noindent and the remaining error\footnote{We notate the remaining error as $\vec{R}$ instead of $\sigma^2$ as in \cite{Andrade2019} since the term can be negative.}, $\vec{R}$, is given by

\begin{equation}
    \label{eq: reconst ct}
    {R}_i = \sum^{M}_{j=L+1}\sum^{M}_{j'=L+1}C_{i, j} \covnv{a_{\mathrm{r}, j}}{a_{\mathrm{r}, j'}} C^\mathrm{T}_{j', i} + 2\sum^{M}_{j=L+1}C_{i, j} \covnv{a_{\mathrm{f}, i}}{a_{\mathrm{r}, j}},
\end{equation}

\noindent where $L$ is the number of fitted modes, $M$ is the total number of modes, $i$ represents the Zernike mode, and $C$ are terms of the cross-talk matrix. Since Zernike polynomials are used in the real-time control of the NAOMI system, we opted to use them in favour of Karhunen-Loève modes. The Zernike basis is not statistically independent. This introduced the additional cross-term, $\cov{a_\mathrm{r}}{a_\mathrm{f}}$, in the remaining error. There was then an additional explicit dependence on the remaining modes. As a result of this dependence, the $\vec{R}$ is larger when compared to statistically independent modes (the K-L modes) \citep{Dai1996}.

\subsection{Turbulence parameter estimation algorithm}

The integrated turbulence parameters, $\vec{p} = \left[ r_0, \mathcal{L}_0 \right]$, were estimated through the minimisation of the norm of the error function

\begin{equation}
    \label{eq: minimisation of p}
    \vec{p} = \arg \min_{\vec{p}}{\{||\vec{\epsilon}||^2\}},\end{equation}

\noindent where

\begin{equation}
    \label{fun: error function}
    \vec{\epsilon} = \var{\vec{b}}{} - \left( \var{\vec{a}}{f}_\mathrm{vk}(\vec{p}) + \vec{R}(\vec{p}) + \vec{\sigma^2_{\mathrm{n}}} \right).
\end{equation}

\noindent This function is a vector in the Zernike modes that includes the analytical models for the von Kármán variances, $\var{a}{}_\mathrm{vk}(\vec{p})$, the remaining error, $\vec{R}(\vec{p})$, and the measurement error, $\vec{\sigma^2_{\mathrm{n}}}$.

A $\chi^2$ approach was employed \citep{bevington2003data} to solve equation (\ref{eq: minimisation of p}), thus expanding on previous work by \cite{Andrade2019}. The $\chi^2$ approach requires knowledge of the squared uncertainty of each data point, in our case we assumed that the uncertainty of each radial order is constant. This constant value, $\vec{\sigma^2_{\var{b}{}}}$, was estimated by taking the variance of the reconstructed modes, $\var{b}{}$, in each radial order.

The uncertainty follows the approximate logarithmic scale of the von Kármán modal variances (c.f. Appendix \ref{annex: Reconstructed variances}). 
Each radial order has a singular theoretical modal variance shared between azimuthal orders, and, as such, any deviation from this variance is a deviation from the mean. This approach has the advantage of casting the minimisation problem as a known $\chi^2$ problem and achieving a homogeneous contribution of all radial orders to the fit without forcing a logarithmic scale, which was the case in previous implementations \citep{Andrade2019}.

The measurement and remaining error contributions to the $\vec{\epsilon}^2$ function are much smaller in magnitude compared to $\var{a}{f}_\mathrm{vk}$. Therefore, we chose to separate the estimation of the former from the global fit performed in equation (\ref{eq: minimisation of p}):

\begin{itemize}
    \item The measurement error was estimated through the use of the autocorrelation of the Zernike coefficients \citep{Fusco2004} (c.f. Appendix \ref{annex: Fusco noise}). This method estimates the measurement error as a vector across Zernike modes. Due to the reconstruction matrix, this is the case for the measurement error expressed in equation (\ref{eq: measurement error}). Finally, this new approach was shown in simulation to perform better than a global fitting approach \citep{Andrade2019} at a signal-to-noise ratio $(\mathrm{S/N}) < 1$, while performing equally as well at a higher $\mathrm{S/N}$ (c.f. Appendix \ref{annex: Fusco noise comparison}). We define the $\mathrm{S/N}$ as
    \begin{equation}
        \label{eq: SNR}
        \mathrm{S/N} = \frac{\var{\vec{s}}{}}{\sigma^2_0}.
    \end{equation}
    
    \item The remaining error analytical expression (c.f. Eq. \ref{eq: reconst ct}) is dependent on $\vec{p}$. Although we could estimate both jointly, we chose to estimate each iteratively \citep{Andrade2019}. An iterative estimation of $\vec{R}(\vec{p})$ was found to be numerically much more convenient with the algorithm converging in a small number of iterations on all cases tested. 
\end{itemize} 

The iterative $\chi^2$ algorithm is then expressed as 

\begin{equation}
    \label{eq: full algorithm}
    \vec{p^k} = \arg \min_{\vec{p}}\sum^{L = 15}_{i=4}\left\{ \frac{ \varnv{b}{i} - f(\vec{p^{k}}) }{\sigma_{\varnv{b}{},i}} \right\}^2,
\end{equation}
\noindent with 
\begin{equation}
    f(\vec{p^{k}}) =
    \begin{cases}
          \left(\varnv{a}{\mathrm{f},i}_\mathrm{vk} \right)(\vec{p^k}) + \sigma^2_{\mathrm{n},i}, & \text{if}\ k=0 \\
          \, \\
          \left(\varnv{a}{\mathrm{f},i}_\mathrm{vk} \right)(\vec{p^k}) + {R}_i(\vec{p^{k-1}}) + \sigma^2_{\mathrm{n},i}, & \text{if}\ k>0
    \end{cases},
\end{equation}

\noindent where $k$ represents the iteration of the algorithm and $i$ represents the i-th Zernike coefficient, thus summing over the terms of the variance vector. 
The tip and tilt modes were removed from the range of the fit (i.e. the minimisation starts at $i=4$) as they are still largely affected by wind shake and telescope vibrations \citep{gluck2017investigations}. While the defocus mode is also affected, we chose to use it in the reconstruction due to the small number of available modes. The maximum reconstructed modal variance included in the fit, $L$, is dictated by the number of modes controlled by the NAOMI system $(L=15)$ \citep{Woillez2019}.

\subsection{Algorithm convergence and sensitivity}

\begin{figure}[ht]
            \resizebox{\hsize}{!}{\includegraphics{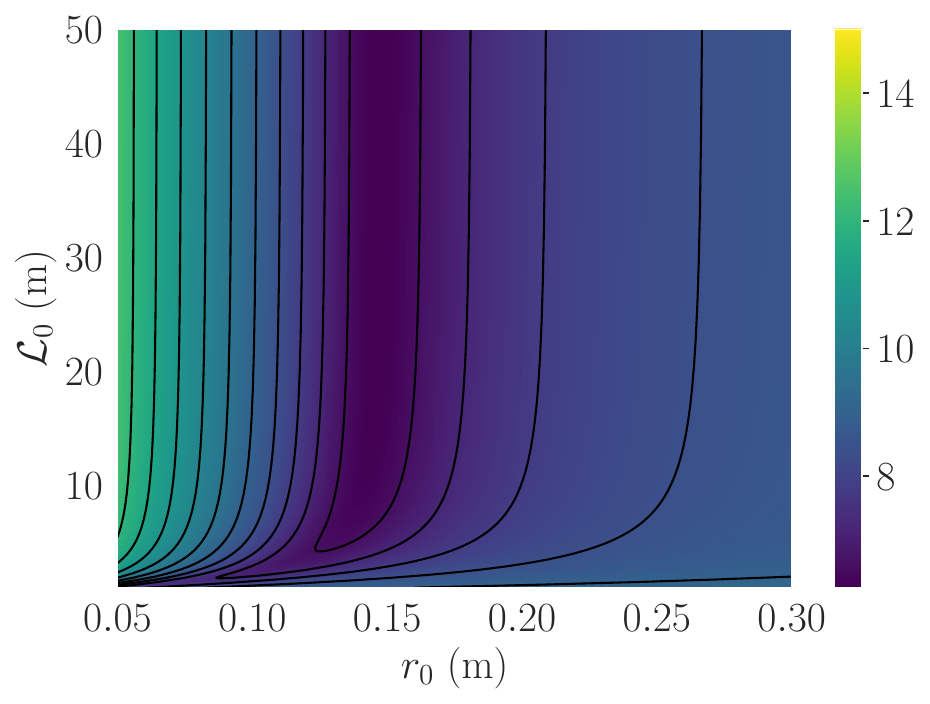}}
            \caption{$\chi^2$ minimisation map for equation (\ref{eq: full algorithm}) using simulated telemetry data. The colour bar represents the chi-squared value in a logarithmic scale.}
            \label{fig: 2D chi squared map}
\end{figure}

The convexity is shown in the $\chi^2$ map of Figure \ref{fig: 2D chi squared map}. The Fried parameter is shown to have a well-defined minimum, whereas the outer scale does not.

This confirms an expected result: the minimum measurable spatial frequency determines the maximum spatial scale detectable by the sensor. In the case of the ATs, the minimum measurable spatial frequency is $ f_\mathrm{m} = 1/D =  \SI{0.556}{\m^{-1}}$, where $D$ is the telescope diameter. As such, NAOMI is insensitive to spatial scales larger than its $D=\SI{1.8}{\m}$. The outer scale has been shown to vary between 
$(1-100)\,\SI{}{\m}$ \citep{Martin2000,Ziad2004}, leading the ATs to be insensitive to most outer scale conditions.

\subsection{Estimation of uncertainty}
\label{Estimation of uncertainty}

The following method estimates the statistical uncertainty of the fit, $u(\mathbf{p})|_\mathrm{stat}$. The statistical uncertainty is added to the systematic uncertainty of the algorithm to serve as a confidence interval for the estimated $r_0$ when the algorithm is applied to on-sky telemetry, where the turbulence parameters are unknown.  

We implemented a Monte Carlo simulation for the estimation of the uncertainty. This methodology uses random sampling to study the significance of the data. The random sample must mirror our data's statistical behaviour. For each radial order, $n$, the random sample, $\var{b}{\mathrm{MC},i}$, follows a normal distribution,

\begin{equation}
    \left[\var{b}{\mathrm{MC},i}\right]_n = \left[\mathcal{N}\left( \overline{\var{b}{i}} ,\sigma_{\var{b}{},i}\right)\right]_n,
\end{equation}  

\noindent where the $\overline{\var{b}{i}}$ is the mean modal variance of the radial order, $n$, for the reconstructed variances, $\var{\vec{b}}{}$, and $\sigma_{\var{b}{},i}$ is the standard deviation around the mean. A new random sample, $\var{\vec{b}}{MC}$, was obtained by sweeping across the modes in Eq. (\ref{eq: full algorithm}) (i.e. $i=4$ to $i=15$). When applying the estimation algorithm to the new sample, $\var{\vec{b}}{\mathrm{MC}}$, we obtained a new set of turbulence parameters from which we evaluated the uncertainty.

The Monte Carlo algorithm generates $H=50$ random samples. The sample size was a good compromise between the computational load and the performance of the estimation. Since the study of the uncertainty is done by a repeated generation of new values of turbulence parameters, we are interested in calculating the error of the mean estimation. From the $H$ turbulence parameters, the uncertainty of the mean is given by the standard deviation of the mean \citep{bevington2003data}. For the Fried parameter, the statistical uncertainty is given by

\begin{equation}
    u({r_0})|_\mathrm{stat} = \frac{1}{H}\left[\sqrt{\sum^H_{i=1}\left(r_{0,i} - \overline{r_0}\right)^2}\right],
\end{equation}

\noindent where $\overline{r_0}$ is the mean of the Fried parameter for the $H$ samples and $r_{0, i}$ is the Fried parameter resulting from the estimation algorithm for each sample.

Section \ref{sec: results} uses the seeing, $\alpha$, as a comparison tool to other independent estimation methods. It is obtained from the $r_0$ through \citep{Fried1965}

\begin{equation}
    \alpha = 0.976\frac{\lambda}{r_0}
\end{equation}

\noindent at a wavelength, $\lambda$, of $\SI{500}{\nm}$. The uncertainty of the seeing, $u(\alpha)$, is obtained directly from the uncertainty of the $r_0$ through error propagation analysis. The $u(\alpha)$ is then given by

\begin{equation}
    u(\alpha) = \frac{\alpha}{r_0}u(r_0),
\end{equation}

\noindent where all variables, excluding the $r_0$, are assumed to be true values with no uncertainty.

\section{Algorithm validation and optimised performance}
\label{subsec: simulation results}

\subsection{Simulation setup}

\begin{table}[ht]
    \caption{Features of the NAOMI AO system.}
    \label{tab:features of NAOMI}
    \centering
    \vspace{0.3 cm}
    \begin{tabular}{c c}
    \hline\hline  
    Feature & \\
    \hline
    Diameter & 1.8 m\\
    Obstruction ratio & 7.7 \%\\
    Loop frequency & 500 - 50 Hz\\
    Pixel scale & 0.375 ''pix$^{-1}$\\
    Sub-apertures & 4x4 \\
    Valid sub-apertures & 12\\
    Total loop delay & 4.6 ms\\
    Corrected modes & 4-15\\
    \hline
    \end{tabular}
\end{table}

\begin{table}[ht]
    \caption{Simulation conditions used in the generation of the phase screens.}
    \label{tab: simulation conditions}
    \centering
    \begin{tabular}{c c}
    \hline\hline  
    Feature & \\
    \hline
    Fried Parameter     & 14.6 cm \\
    Outer scale & 18.9 m \\
    Type of phase screens & Correlated\\
    \# layers & 9 \\
    \hline
    \end{tabular}
\end{table}

The estimation algorithm was optimised entirely in \textit{python} using the Object Oriented Python Adaptive Optics (OOPAO) package \citep{Heritier2023}, which inherits the structure and functionality of its Matlab precursor OOMAO \citep{conan14}. The generation of von Kármán atmospheric phase screens was done using the algorithm proposed by \cite{assemat2006method} using the \textit{AOtools} package \citep{townson2019aotools}.

The atmospheric parameters were kept constant throughout the simulation due to the assumed stationarity. Turbulence parameters vary in time, but we admit that they are stationary in the range of 1-2 minutes of acquisition \citep{Ziad2004}. 

\begin{figure}[ht]
            \resizebox{\hsize}{!}{\includegraphics{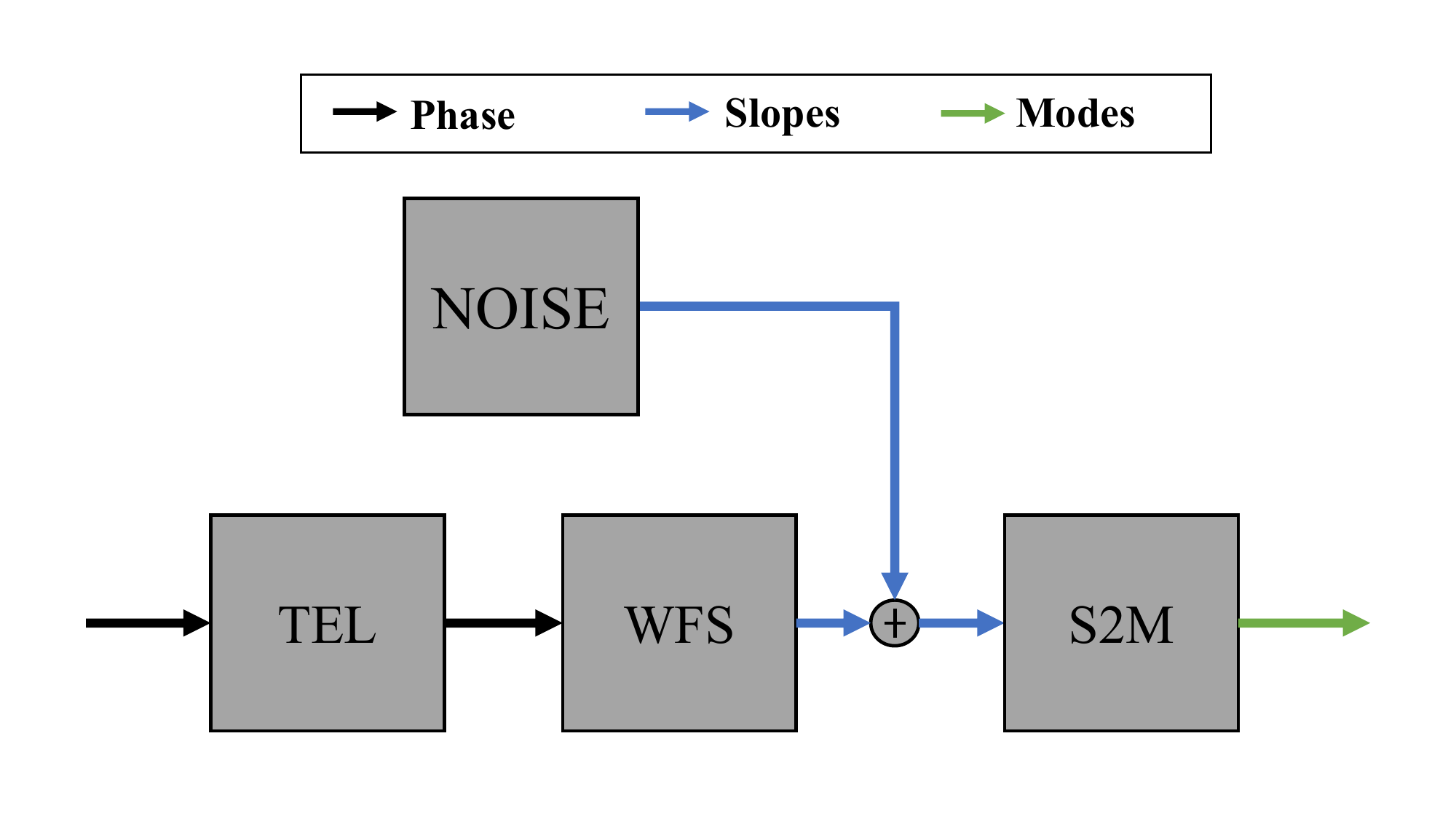}}
            \caption{Open-loop block diagram for the simulation. The phase screens were propagated through the telescope object, TEL, and were converted into a slope measurement by the wavefront sensor model, WFS. Measurement error, NOISE, was added to the slopes. The slopes were then converted into Zernike coefficients by matrix multiplication with the S2M matrix.}
            \label{fig: block diagram}
\end{figure}

The simulation was performed in an open loop, implemented by propagating the generated phase screens through a telescope object and a diffractive model of the WFS with a lenslet field of view of $\SI{7.33}{\arcsecond}$ to ensure linearity (Figure \ref{fig: block diagram}). The slopes obtained from the SH-WFS were converted into Zernike modal coefficients using a synthetic control matrix. Finally, the variances of the reconstructed coefficients were introduced to the estimation algorithm to obtain the $r_0$. The simulation was done in open-loop since the von Kármán model exists for uncorrected variances.

The telescope and WFS were configured per the prescription in \cite{Woillez2019} and are summarised in Table \ref{tab:features of NAOMI}. The propagating phase screens are correlated and reflect the typical turbulence parameters of the Paranal Observatory (Table \ref{tab: simulation conditions}). Nine layers were used for the atmospheric model with the wind speed and altitude profiles following the values presented by \cite{Masciadri2013}. Screens were generated at a $\SI{500}{\Hz}$ loop frequency. SH-WFS measurements feature a Gaussian additive white noise with a known variance ($\sigma = [10^{-6}-10^{-7}] $, c.f. \ref{eq: noise def}). In the simulation, this corresponds to a 0.1 - 25 S/N regime. The average noise magnitude was taken from a $N=50$ telemetry data sample representative of the entire set.

The telemetry samples already contain a gradient matrix, $\mat{G}_\mathrm{system}$, obtained from the inverted control matrix that accounts for the first 14 controlled modes. A much higher number of modes is needed for the estimation of the remaining error (c.f. equation (\ref{eq: reconst ct})). The estimation algorithm used a synthetic matrix, $\mat{G}_\mathrm{synthetic}$, truncated between the second and $13^\mathrm{th}$ radial order (i.e. 90 modes) to calculate these terms. The synthetic matrix was obtained by registering the response of the Shack-Hartmann model to individual Zernike modes introduced at the telescope pupil. The matrix was validated against the $\mat{G}_\mathrm{system}$ available in the telemetry samples and extended to additional modes once validation was achieved. The synthetic and system matrices were compared, and a median error of 1.3\% was found between the two. This mismatch adds systematic error to our estimation algorithm.  

The convergence bias of the algorithm was evaluated by comparing the turbulence parameters used to generate the phase screens, $r_\mathrm{0,screen}$, with the turbulence parameters obtained from the minimisation of the algorithm, $r_\mathrm{0,fit}$. For example, the per-cent relative difference, 

\begin{equation}
    \widetilde{\Delta r_0} = \frac{|r_\mathrm{0,fit} - r_\mathrm{0,screen}|}{r_\mathrm{0,screen}}(\%),
\end{equation}

\noindent between the estimation of the algorithm and screen value gives us the accuracy of the estimated $r_{0}$.

\subsection{Algorithm parameter optimisation}

\begin{figure}[ht]
            \centering
            \resizebox{\hsize}{!}{\includegraphics{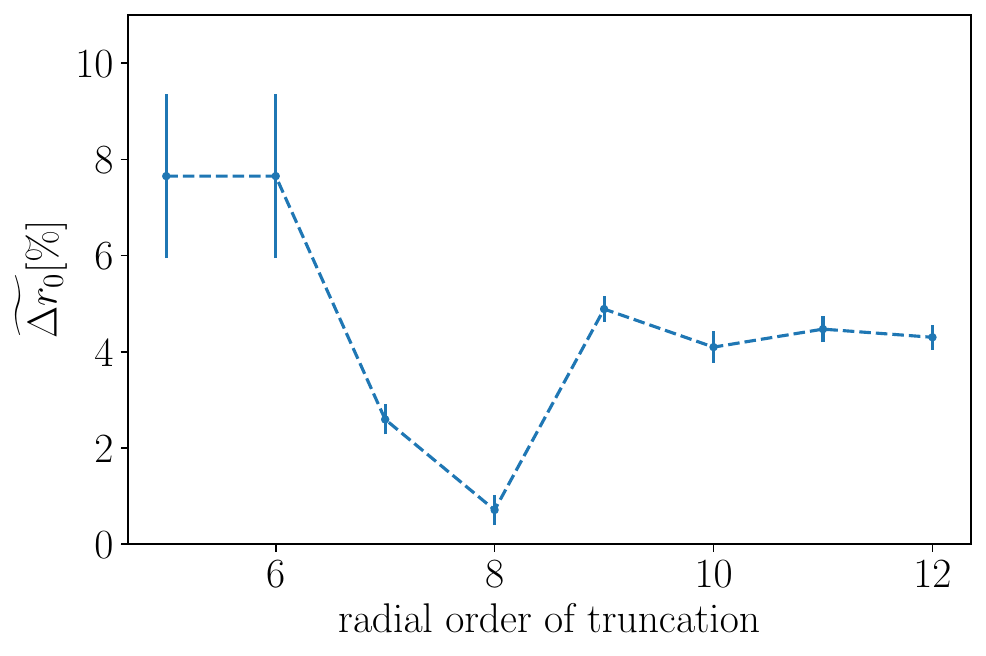}}
            \caption{$\widetilde{\Delta r_0}$ as a function of the maximum radial order included in the estimation of the remaining error. The upper limit was allowed to vary by changing the maximum radial order included in the $\mat{G_\mathrm{r}}$ matrix. A time horizon of $\SI{20}{\s}$ was chosen for the generation of the curve. Error bars present a $\sigma$ deviation to the sample's mean $\widetilde{\Delta r_0}$.}
            \label{fig: evol with Gr}
\end{figure}

\begin{table}[ht]
    \caption{Optimised parameters used in the iterative algorithm.}
    \label{tab: estimation features}
    \centering
    \begin{tabular}{c c}
    \hline\hline  
    Feature & \\
    \hline
    Maximum radial order of fit     &  4 \\
    Maximum radial order of remaining error & 8 \\
    Minimum \# of iterations & 2 \\
    \hline
    \end{tabular}
\end{table}

Various fitting features of the iterative algorithm were tested and optimised; their optimised values are presented in Table \ref{tab: estimation features}. From the tested features: 
\begin{itemize}
    \item The maximum radial order of the fit is the upper limit of fitting, $L$, of equation (\ref{eq: full algorithm}). The ideal number of modes was found to meet the condition of inclusion of all controlled modes by NAOMI ($L=15$);
    \item The maximum radial order of the remaining error corresponds to the maximum number of modes included in the calculation of the cross-talk matrix. The matrix was constructed from the corrected terms, $\mat{H^+}$, and the higher non-corrected terms, $\mat{G_\mathrm{r}}$. Letting the $\mat{G_\mathrm{r}}$ matrix vary in size makes it possible to control the number of radial orders present in the estimation of the remaining error. Thus the estimation of turbulence parameters can be made to vary as a function of the maximum radial order included in the calculation of $\vec{R}$. We note that $\widetilde{\Delta r_0}$ was found to stabilise once the error estimation included the eighth radial order (Figure \ref{fig: evol with Gr}). The optimal radial order does not coincide with the maximum available radial order. This is attributed to a mismatch between the models used for the simulation of the optical system and the one used for the estimation of $f(\vec{p})$; 
    \item The minimum number of iterations, $k$, for which the estimation algorithm converges to a set of fitting parameters, $\vec{p^k}$, was shown to be small ($k = 2$). Once the threshold of $k=2$ iterations is crossed, only sub-per-cent changes to the $\widetilde{\Delta r_0}$ are registered;
    \item The ideal starting parameters are obtained from the analysis of Figure \ref{fig: 2D chi squared map}. An initial $\vec{p}$ with a small value of $r_0$ is ideal for the minimisation, as the derivative of the $\chi^2$ is larger for small values of the Fried parameter.

\end{itemize}

\subsection{Truncation of the time series}

\begin{figure}[ht]
    \resizebox{\hsize}{!}{\includegraphics{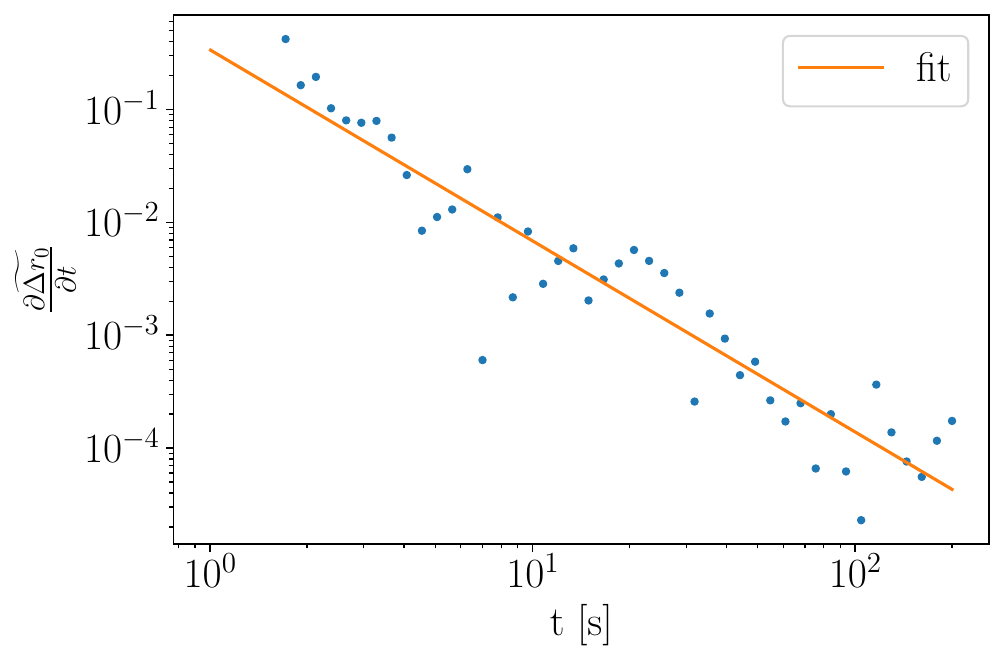}}
    \caption{Evolution of the convergence of the algorithm with time. The convergence of the algorithm was studied by monitoring the derivative of the turbulence parameter, ${\partial \widetilde{\Delta r_0}}/{\partial t}$, estimate as a function of the time horizon, $t$.}
    \label{fig: convergence of r0}
\end{figure}

A large portion of our samples ($\SI{88.50}{\%}$) has a $\SI{20}{\second}$ time horizon. We are interested in analysing the effects of truncation of the time series on the temporal convergence of the algorithm. The temporal evolution of the estimated Fried parameter,
    
\begin{equation}
    \frac{\partial \widetilde{\Delta r_0}}{\partial t} \approx \frac{(r_{0,n+1} - r_{0,n})/r_\mathrm{0,screen}}{\Delta t},
\end{equation}
    
\noindent is represented in Figure \ref{fig: convergence of r0}. The temporal evolution is studied by comparing the change in parameter estimation from time horizon $t$ to $t + \Delta t$. The loop frequency of the system gives the time increment $\Delta t$. The temporal evolution is modelled as the power-law decay,

\begin{equation}
    \label{eq: power law}
    \frac{\partial \widetilde{\Delta r_0}}{\partial t} = C + \left(\frac{\tau}{t}\right)^\alpha.
\end{equation}

\noindent The temporal evolution for a $\SI{20}{\second}$ time horizon was retrieved from the fit of the model and was found to add an uncertainty of $\SI{0.2(2)}{\%}$. The error of the uncertainty was calculated from the variance of the fitted parameters of the fit.

The convergence bias is assumed to be independent of the mismatch between synthetic and system matrices. The total systematic error is then the square sum of the two errors ($u(r_0)|_\mathrm{sys} = \SI{1.2}{\%}$).

\subsection{Optimised algorithm performance}

\begin{figure}[ht]
            
            \resizebox{\hsize}{!}{\includegraphics{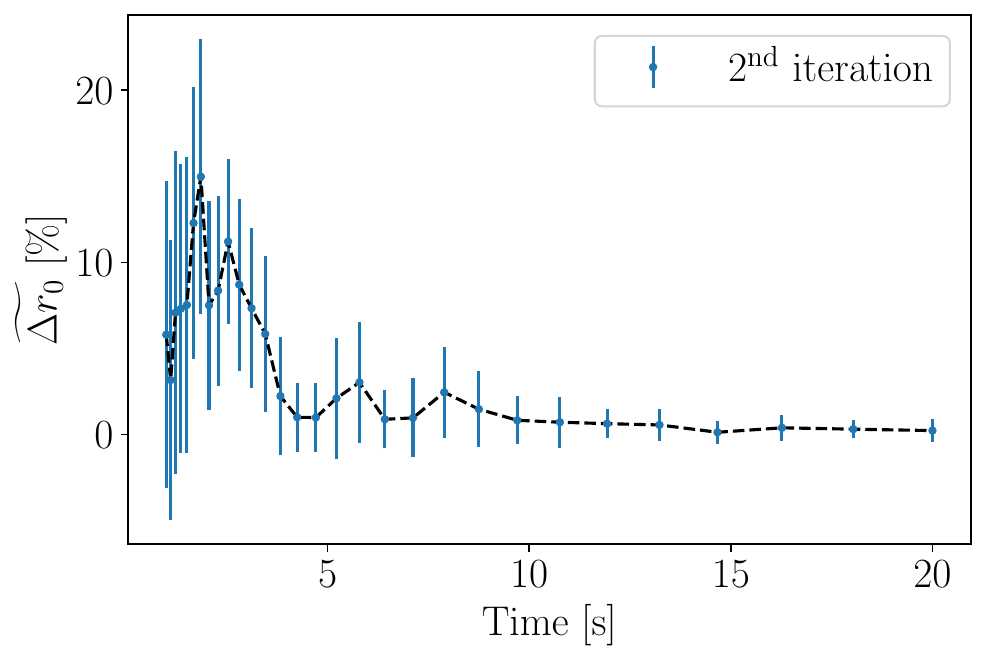}}
            \caption{Results for the simulated data ($N=17$ samples). The second iteration estimation is represented. A maximum time horizon of $\SI{20}{\s}$ is presented. Error bars represent a $3\sigma$ deviation to the sample's mean $\widetilde{\Delta r_0}$.}
            \label{fig: simulation results}
\end{figure}

Figure \ref{fig: simulation results} indicates the algorithm's performance in estimating the Fried parameter as a function of the time horizon of the sample. The effects of the remaining error estimation were tested by comparing the $k = 0$ and $k = 2$ iteration of the algorithm. The algorithm's $k=0$ iteration does not estimate the remaining error, while any iteration above the second was shown to be unnecessary. The optimised algorithm shows that the correction of the remaining error is necessary. Between the two iterations of the algorithm, an improvement in convergence bias from $\SI{17}{\%}$ to a sub-per cent is verified.

The algorithm's performance answers the feasibility question posed in Section \ref{sec: intro}. Sub-per-cent convergence bias estimations can be obtained in intervals of $\SI{20}{\second}$ non-overlapping telemetry samples once the remaining error estimation is included in the fit for a $4\times4$ Shack-Hartmann sensor.

\section{Results for NAOMI on-sky telemetry} 
\label{sec: results}

\subsection{Dataset analysis}

\begin{figure}[ht]
            \resizebox{\hsize}{!}{\includegraphics{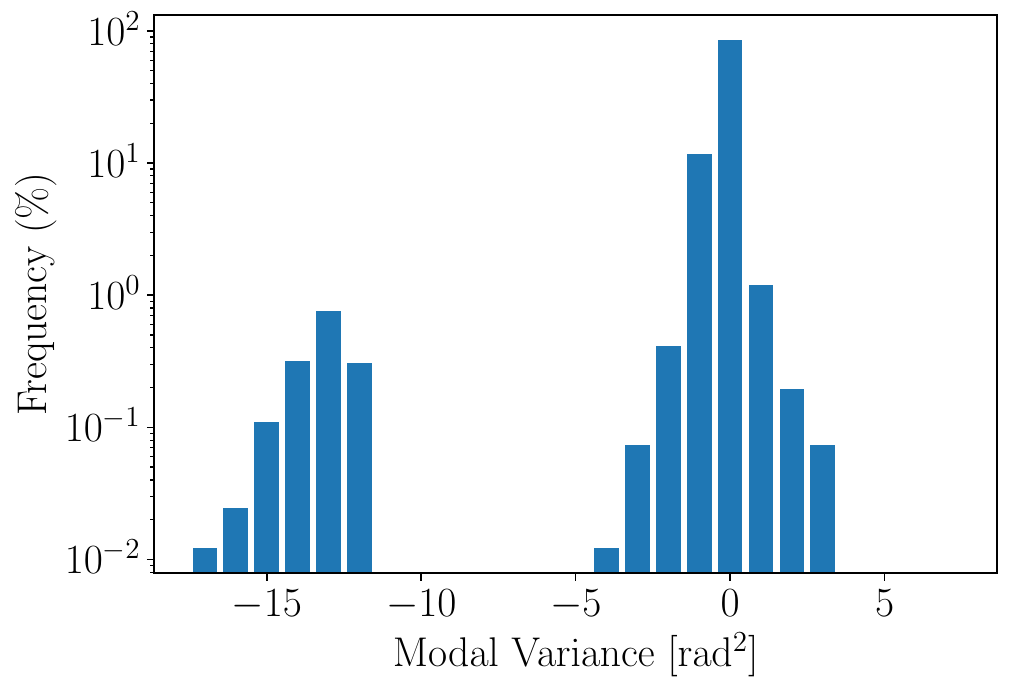}}
            \caption{Logarithmic histogram representation of the variance of the defocus mode for our data sample. Data points below an exponent of E$-10$ are considered to be empty datasets and have been removed from our sample.}
            \label{fig: temp Zern coeff outlier}
\end{figure}

\begin{table}[ht]
\caption{Features of the NAOMI telemetry dataset.}
    \centering
    \begin{tabular}{c c}
    \hline\hline  
    Feature & \\
    \hline
    \# of samples     & 8170 \\
    500Hz loop frequency    & $\SI{92.44}{\%}$ \\
    Open/Closed Loop data     & (10, 90)$\SI{}{\%}$ \\
    Date range     &  2018/11 - 2020/12 \\
    \# of SH-WFS sensors     &  4 \\
    $\SI{20}{\s}$ time horizon & 88.50 \% \\
    \hline
    \end{tabular}
    \label{tab: features of cluster}
\end{table}

The on-sky data were packaged following the telemetry data format proposed by \cite{Gomes2022}. These data were curated following several criteria summarised in Table \ref{tab: features of cluster}:

\begin{itemize}
    \item Data with a $\SI{500}{\Hz}$ loop rate were selected. This loop ratio ensures a sufficient Strehl ratio in the samples. Lower loop frequencies are only active for high-R magnitude AO reference star samples with low H- and K-band Strehl ratios \citep{Woillez2019};
     \item A minimum time horizon of $\SI{20}{\s}$ was selected. We note that $\SI{88.50}{\%}$ of the available data satisfy this condition;
     \item $N = 166$ telemetry samples possess unviable data (c.f. Figure \ref{fig: temp Zern coeff outlier}). We consider samples with a defocus $\var{b}{}$ below a $\SI{1e-10}{\radian^2}$ threshold to be incorrectly stored and remove them from the sample. 
\end{itemize}

\begin{figure}[ht]
             \resizebox{\hsize}{!}{\includegraphics{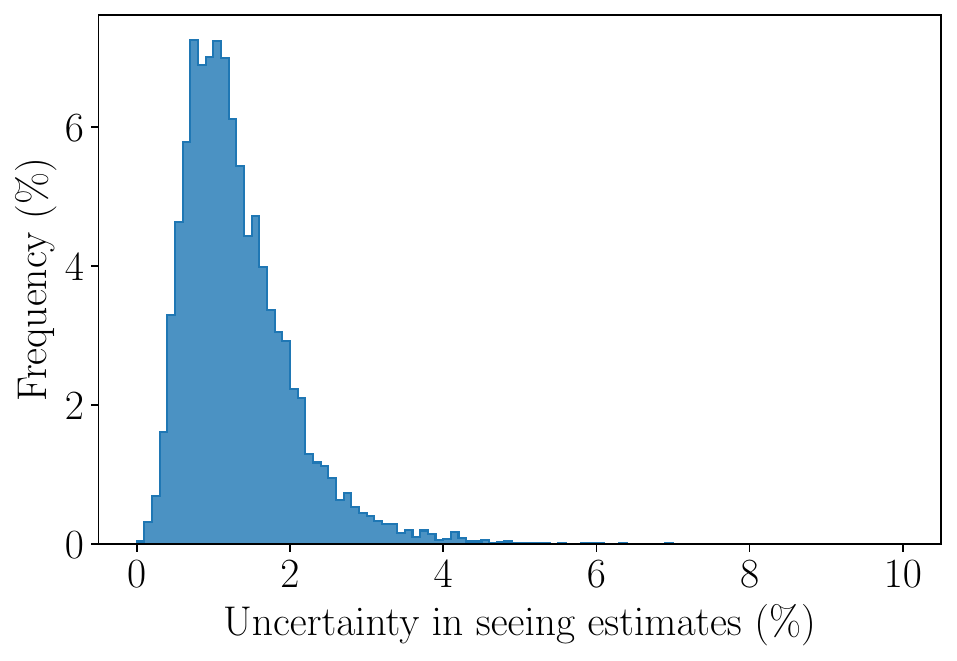}}
             \caption{Uncertainty of seeing estimates. The were obtained from applying the Monte Carlo method developed in Section \ref{sec: methods} to the full NAOMI AO telemetry dataset before removing outliers.} 
             \label{fig: uncertainty estimates}
\end{figure}

We ensure the correct confidence levels of the algorithm through an analysis of the uncertainty of the estimation. The per-cent statistical uncertainty of the algorithm estimation is shown in Figure \ref{fig: uncertainty estimates}. The median statistical uncertainty is 1.2\%. High uncertainty points were removed from the sample. We chose $\SI{10}{\%}$ as a threshold for acceptance. The $N = 73$ data points (less than 1\%) were removed from the data sample this way. The total error of the estimation must include the systematic error. As such, the systematic error was added to the median uncertainty to obtain the median confidence level of ${(\alpha \pm 1.2|_\mathrm{stat} \% \pm 1.2|_\mathrm{sys} \%)}$.

\subsection{Agreement between ATs}

\begin{table}[ht]
    \caption{Comparison of ATs' seeing estimations.}
    \label{tab: comp SH}
    \centering
    \begin{tabular}{c c}
    \hline\hline
    Data     & Mean  \\
    \hline
    $\mathrm{AT}_{1} - \mathrm{AT}_{2}$     &  \SI{0.011}{\arcsecond} \\
    $\mathrm{AT}_{1} - \mathrm{AT}_{3}$    & \SI{-0.000}{\arcsecond}\\
    $\mathrm{AT}_{1} - \mathrm{AT}_{4} $    & \SI{-0.020}{\arcsecond} \\
    $\mathrm{AT}_{2} - \mathrm{AT}_{3} $    & \SI{-0.013}{\arcsecond} \\
    $\mathrm{AT}_{2} - \mathrm{AT}_{4} $   &  \SI{0.010}{\arcsecond} \\
    $\mathrm{AT}_{3} - \mathrm{AT}_{4} $    & \SI{-0.012}{\arcsecond}  \\
    \hline
    \end{tabular}
\end{table}

Samples are gathered from the four ATs of the VLTI. These are spread across the observatory and monitor the same star. The four telescopes provide concurrent AO telemetry samples for the same observation. We can sample the spatial profile of the Paranal Observatory through the former with up to four concurrent seeing estimations. Table \ref{tab: comp SH} summarises the difference in seeing estimates for concurrent telemetry samples by presenting the mean difference between telescopes. The error shown is the error of the mean for the data points. From Table \ref{tab: comp SH}, we conclude that:

\begin{itemize}
    \item the average difference between telescopes is $\SI{0.011}{\arcsecond}$. This difference accounts for a \SI{1.4}{\%} difference in average seeing across ATs. Since the median statistical uncertainty is \SI{1.2}{\%}, and the systematic uncertainty is \SI{1.2}{\%} for a 20 second time horizon, this points to an absence of a spatial profile of the seeing across Paranal within our uncertainty.
    
    \item on average, the error around the mean is $\SI{0.005}{\arcsecond}$, which results in a difference of $\SI{0.7}{\%}$ around the average seeing difference. The dispersion around the mean is small, demonstrating the robustness of the estimates.
    
\end{itemize}

\subsection{Comparison with DIMM}

\begin{figure}[ht]
    \resizebox{\hsize}{!}{\includegraphics{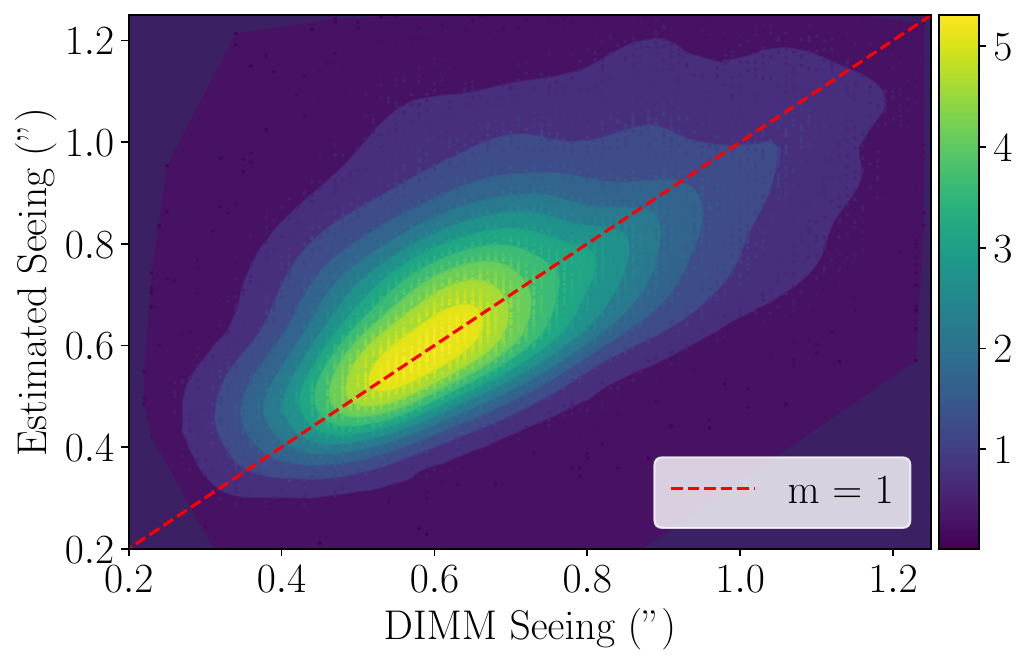}}
    \caption{Relationship between algorithm estimations and DIMM data. The Pearson correlation coefficient is 0.62 between the two datasets. The colour bar represents the density of points of a particular ($\alpha_\mathrm{DIMM},\alpha_\mathrm{fit}$) set.}
    \label{fig: relationship with DIMM}
\end{figure}

\begin{figure}[ht]
    \resizebox{\hsize}{!}{\includegraphics{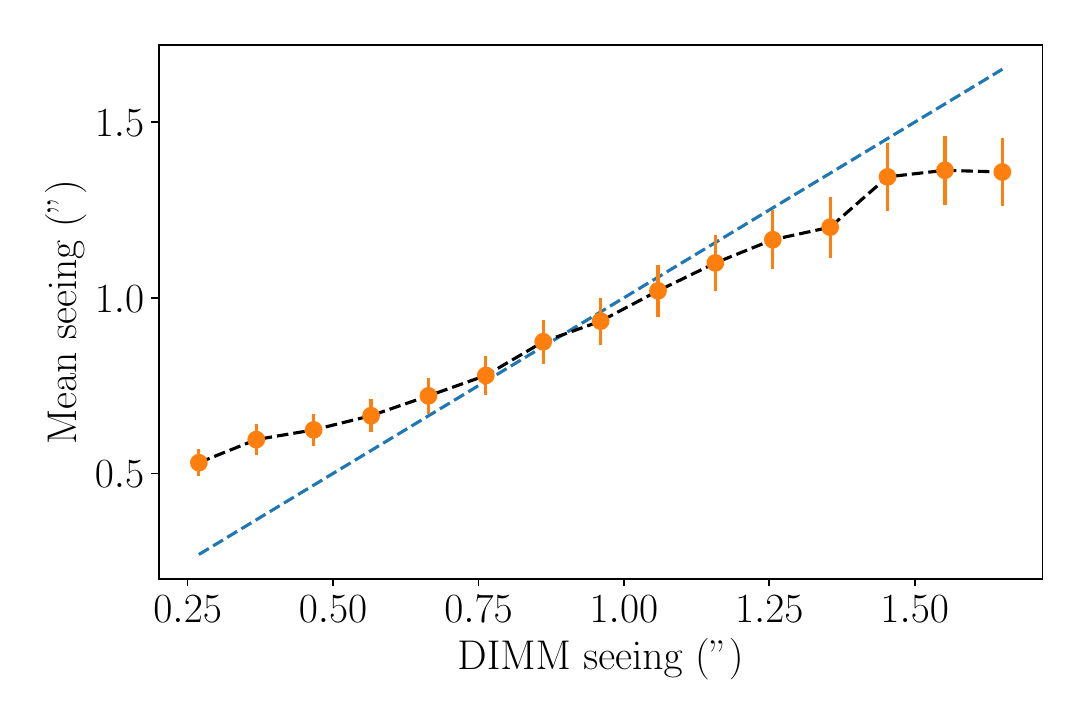}}
    \caption{Algorithm and DIMM seeing relationship trend. Points represent the mean seeing estimation of the algorithm when compared to the DIMM estimates for the same telemetry sample; the error bars present a 3$\sigma$ confidence level of the mean individual estimation.}
    \label{fig: binned seeing relation}
\end{figure}

Differential image motion monitors (DIMM) are implemented at Paranal and give regular seeing estimates \citep{ESO_PAPER}, which are used in this section for comparison to our estimates. The relationship between the two estimates is represented in Figures \ref{fig: relationship with DIMM} and \ref{fig: binned seeing relation}. A linear trend between the two estimates is observed up to a limit of 1.5 arc seconds, as illustrated in Figure \ref{fig: binned seeing relation}. This limit was chosen as a cutoff threshold for the analysis. Within the selected range, the DIMM and our seeing estimates agree with a Pearson correlation coefficient of 0.62. On average, a 5\% larger seeing is estimated ($\overline{\alpha} = \SI{0.76}{\arcsecond}$). 

Following \cite{Masciadri2023} DIMM comparisons, we introduced bias and root-mean-square error ($\mathrm{RMSE}$) defined as

\begin{equation}
    \mathrm{BIAS} = \sum^N_{i=1}\frac{(Y_i - X_i)}{N}
\end{equation}

\noindent and

\begin{equation}
    \mathrm{RMSE} = \sqrt{\sum^N_{i=1}\frac{(Y_i - X_i)^2}{N}},
\end{equation}

\noindent where $N$ is the number of telemetry samples, $X_i$ is the observation of MASS-DIMM, and $Y_i$ is the estimation of our algorithm. We obtained $\mathrm{BIAS} = \SI{0.03}{\arcsecond}$ and $\mathrm{RMSE} = \SI{0.22}{\arcsecond}$. These results are consistent with those reported in \cite{Masciadri2023} for the comparison of values between Stereo-SCIDAR and MASS-DIMM, which are $\mathrm{BIAS} = \SI{0.08}{\arcsecond}$ and $\mathrm{RMSE} = \SI{0.25}{\arcsecond}$, respectively. 

Consequently, these findings follow a larger trend of non-consensual seeing estimates between the MASS-DIMM system and other turbulence estimators \citep{Butterley2020, Osborn2018}. The ATs and the DIMM tower are located at different positions in the observatory, with the ATs occupying a central position while DIMM is installed at the edge of the observatory. The dome turbulence is an additional component in the telemetry data. The dome turbulence has been shown to play an important role in the total value of the seeing in larger telescopes than the ATs \citep{Munro2023,Salmon2009}. DIMM uses an open telescope design \citep{Tokovinin2007}, which does not share a dome with the ATs. Coupling these factors and discarding systematic effects between the DIMM and NAOMI turbulence measurements, the local turbulence at the ATs seems to be larger than the one seen at the edge of the observatory.

\section{Conclusions} 
\label{sec: conclusion}

This paper aimed to determine whether low-order ($4\times 4$) Shack-Hartmann wavefront sensor telemetry could be used to estimate integrated turbulence parameters. To come to a conclusion, an iterative estimation algorithm was first validated on simulated data and then applied to telemetry data produced from the operation of the AO system of the ATs of the VLT, NAOMI.

The estimation algorithm was optimised using a simulation of the NAOMI system. Simulation results show the ideal number of iterations, the minimum time horizon, and the number of modes to be included in the fitted variances and remaining error estimation. The $\chi^2$ map of the algorithm was shown to converge to the expected value without bias, and through its analysis, the ideal initial guess for our algorithm was established. The correction of the remaining and measurement error was shown to be necessary for the correct estimation of the Fried parameter. The uncorrected variances were shown to incorrectly estimate the Fried parameter, with an error of 17\%. By compensating for the remaining error, we were able to achieve sub-per-cent convergence bias in the estimation of the parameter. 

Following optimisation, the algorithm was applied to real on-sky data curated to remove undesirable samples totalling less than 1\%. The uncertainty of the estimation of the seeing was first assessed. A median uncertainty of 1.2\% was obtained.

A comparison of the seeing estimates between concurrent telemetry samples across the four available ATs showed consistency in our estimates. On average, all telescopes estimated the same seeing conditions for the same observation within the measurement uncertainty of \SI{0.005}{\arcsecond}. The seeing was shown to be invariant across ATs. The average difference in seeing between telescopes is $\SI{0.011}{\arcsecond}$. This difference is within the median algorithm uncertainty. 

Our algorithm was compared to the DIMM seeing estimates. Disagreement between the two was found, with a correlation coefficient of 0.62 being determined. Our algorithm estimated $\SI{5}{\%}$ larger seeing conditions. 

It was shown that the Fried parameter could be estimated from a low-order Shack-Hartmann system in a \SI{1.8}{m} diameter telescope but not the outer scale. The next immediate step is to apply the algorithm to a higher-order system in a larger-diameter telescope. Larger telescope diameters imply a larger sensitivity to the outer scale. On the other hand, higher-order systems translate into less contamination of lower orders (more sensitive to the outer scale) via aliasing and crosstalk. The ideal test bench is the CIAO system installed in the Unit Telescopes.

\begin{acknowledgements}
The research leading to these results has received funding from the European Union’s Horizon 2020 research and innovation programme under grant agreement 730890 (OPTICON) and 101004719 (OPTICON RadioNET Pilot), UIDB/00099/2020\_UIPD/00099/2020 - funded by national funds through the FCT/MCTES (PIDDAC), and SFRH/BSAB/142940/2018.

The data used in this work can be obtained in the \href{https://archive.eso.org/wdb/wdb/eso/eso_archive_main/query?prog_id=60.A-9278(D)&max_rows_returned=100}{ESO archive}.

The estimation methods discussed in Section \ref{sec: methods} are made available to the community in the Turbulence Estimation Library (Turlib) Python package \citep{turlib}. We make available the documentation of the code at our \href{https://turlib.readthedocs.io/en/latest/modules.html}{Read the Docs page}. We also provide case scripts for the estimation of turbulence parameters from simulated and on-sky telemetry data. These scripts employ the \textit{aotpy} FITS reading data package \citep{Gomes2022}. 
We acknowledge and thank discussions of the telemetry data standard with Tiago Gomes.

\end{acknowledgements}

\bibliographystyle{aa}
\bibliography{References}

\begin{thebibliography}{55}
\expandafter\ifx\csname natexlab\endcsname\relax\def\natexlab#1{#1}\fi

\bibitem[{{Andrade} {et~al.}(2019){Andrade}, {Garcia}, {Correia}, {Kolb}, \&
  {Carvalho}}]{Andrade2019}
{Andrade}, P.~P., {Garcia}, P. J.~V., {Correia}, C.~M., {Kolb}, J., \&
  {Carvalho}, M.~I. 2019, \mnras, 483, 1192

\bibitem[{Ass{\'e}mat {et~al.}(2006)Ass{\'e}mat, Wilson, \&
  Gendron}]{assemat2006method}
Ass{\'e}mat, F., Wilson, R.~W., \& Gendron, E. 2006, Optics express, 14, 988

\bibitem[{{Avila}(2021)}]{Avila2021}
{Avila}, R. 2021, \mnras, 507, L11

\bibitem[{{Beltramo-Martin} {et~al.}(2019){Beltramo-Martin}, {Correia},
  {Ragland}, {Jolissaint}, {Neichel}, {Fusco}, \& {Wizinowich}}]{Martin2019}
{Beltramo-Martin}, O., {Correia}, C.~M., {Ragland}, S., {et~al.} 2019, \mnras,
  487, 5450

\bibitem[{Bevington \& Robinson(2003)}]{bevington2003data}
Bevington, P.~R. \& Robinson, D.~K. 2003, McGraw-Hill, New York

\bibitem[{{Butterley} {et~al.}(2020){Butterley}, {Wilson}, {Sarazin},
  {Dubbeldam}, {Osborn}, \& {Clark}}]{Butterley2020}
{Butterley}, T., {Wilson}, R.~W., {Sarazin}, M., {et~al.} 2020, \mnras, 492,
  934

\bibitem[{Cantalloube {et~al.}(2020)Cantalloube, Milli, B{\"o}hm, Crewell,
  Navarrete, Rehfeld, Sarazin, \& Sommani}]{cantalloube2020}
Cantalloube, F., Milli, J., B{\"o}hm, C., {et~al.} 2020, Nature Astronomy, 4,
  826

\bibitem[{Conan \& Correia(2014)}]{conan14}
Conan, R. \& Correia, C. 2014, in Proc. of the SPIE, Vol. 9148,
  91486C--91486C--17

\bibitem[{{Dai}(1996)}]{Dai1996}
{Dai}, G.-M. 1996, Journal of the Optical Society of America A, 13, 1218

\bibitem[{{Doelman}(2020)}]{Doelman2020}
{Doelman}, N. 2020, \mnras, 491, 4719

\bibitem[{{F{\'e}tick} {et~al.}(2018){F{\'e}tick}, {Neichel}, {Mugnier},
  {Montmerle-Bonnefois}, \& {Fusco}}]{Fetick2018}
{F{\'e}tick}, R. J.~L., {Neichel}, B., {Mugnier}, L.~M., {Montmerle-Bonnefois},
  A., \& {Fusco}, T. 2018, \mnras, 481, 5210

\bibitem[{Fried(1965)}]{Fried1965}
Fried, D.~L. 1965, JOSA, Vol. 55, Issue 11, pp. 1427-1435, 55, 1427

\bibitem[{Fusco {et~al.}(2004)Fusco, Rousset, Rabaud, Gendron, Mouillet,
  Lacombe, Zins, Madec, Lagrange, Charton, Rouan, Hubin, \&
  Ageorges}]{Fusco2004}
Fusco, T., Rousset, G., Rabaud, D., {et~al.} 2004, Journal of Optics A: Pure
  and Applied Optics, 6, 585

\bibitem[{Gl{\"u}ck {et~al.}(2017)Gl{\"u}ck, Pott, \&
  Sawodny}]{gluck2017investigations}
Gl{\"u}ck, M., Pott, J.-U., \& Sawodny, O. 2017, Publications of the
  Astronomical Society of the Pacific, 129, 065001

\bibitem[{{Gomes} {et~al.}(2022){Gomes}, {Correia}, {Bardou},
  {Beltramo-Martin}, {Fusco}, {Kulcs{\'a}r}, {Morris}, {Moruj{\~a}o},
  {Neichel}, {Osborn}, \& {Garcia}}]{Gomes2022}
{Gomes}, T., {Correia}, C., {Bardou}, L., {et~al.} 2022, in Society of
  Photo-Optical Instrumentation Engineers (SPIE) Conference Series, Vol. 12185,
  Adaptive Optics Systems VIII, ed. L.~{Schreiber}, D.~{Schmidt}, \&
  E.~{Vernet}, 121850H

\bibitem[{{Griffiths} {et~al.}(2023){Griffiths}, {Osborn}, {Farley},
  {Butterley}, {Townson}, \& {Wilson}}]{Griffiths2023}
{Griffiths}, R., {Osborn}, J., {Farley}, O., {et~al.} 2023, Optics Express, 31,
  6730

\bibitem[{{Guesalaga} {et~al.}(2021){Guesalaga}, {Ayanc{\'a}n}, {Sarazin},
  {Wilson}, {Perera}, \& {Le Louarn}}]{Guesalaga2021}
{Guesalaga}, A., {Ayanc{\'a}n}, B., {Sarazin}, M., {et~al.} 2021, \mnras, 501,
  3030

\bibitem[{Hardy(1998)}]{hardy1998adaptive}
Hardy, J.~W. 1998, Adaptive optics for astronomical telescopes, Vol.~16 (Oxford
  University Press on Demand)

\bibitem[{Heritier(2023)}]{Heritier2023}
Heritier, C. 2023, aO4ELT7 conference proceedings

\bibitem[{{Herrmann}(1981)}]{Herrmann1981}
{Herrmann}, J. 1981, Journal of the Optical Society of America (1917-1983), 71,
  989

\bibitem[{{Jolissaint} {et~al.}(2018){Jolissaint}, {Ragland}, {Christou}, \&
  {Wizinowich}}]{Jolissaint2018}
{Jolissaint}, L., {Ragland}, S., {Christou}, J., \& {Wizinowich}, P. 2018, \ao,
  57, 7837

\bibitem[{Karman(1948)}]{Karman1948}
Karman, T.~V. 1948, Proceedings of the National Academy of Sciences of the
  United States of America, 34, 530

\bibitem[{{Kolmogorov}(1941)}]{Kolmogorov1941}
{Kolmogorov}, A.~N. 1941, Proceedings of the Royal Society of London Series A,
  434, 9

\bibitem[{{Kornilov} {et~al.}(2007){Kornilov}, {Tokovinin}, {Shatsky},
  {Voziakova}, {Potanin}, \& {Safonov}}]{Tokovinin2007}
{Kornilov}, V., {Tokovinin}, A., {Shatsky}, N., {et~al.} 2007, \mnras, 382,
  1268

\bibitem[{{Lacour} {et~al.}(2019){Lacour}, {Dembet}, {Abuter}, {F{\'e}dou},
  {Perrin}, {Choquet}, {Pfuhl}, {Eisenhauer}, {Woillez}, {Cassaing},
  {Wieprecht}, {Ott}, {Wiezorrek}, {Tristram}, {Wolff}, {Ram{\'\i}rez},
  {Haubois}, {Perraut}, {Straubmeier}, {Brandner}, \& {Amorim}}]{GRAVITY2019}
{Lacour}, S., {Dembet}, R., {Abuter}, R., {et~al.} 2019, \aap, 624, A99

\bibitem[{Lai {et~al.}(2019)Lai, Withington, Laugier, \& Chun}]{Lai2019}
Lai, O., Withington, J.~K., Laugier, R., \& Chun, M. 2019, Monthly Notices of
  the Royal Astronomical Society, 484, 5568

\bibitem[{{Liu} {et~al.}(2015){Liu}, {Giordano}, {Yao}, {Vernin}, {Chadid},
  {Wang}, {Yin}, \& {Wang}}]{Liu2015}
{Liu}, L.~Y., {Giordano}, C., {Yao}, Y.~Q., {et~al.} 2015, \mnras, 451, 3299

\bibitem[{{Martin} {et~al.}(2000){Martin}, {Conan}, {Tokovinin}, {Ziad},
  {Trinquet}, {Borgnino}, {Agabi}, \& {Sarazin}}]{Martin2000}
{Martin}, F., {Conan}, R., {Tokovinin}, A., {et~al.} 2000, \aaps, 144, 39

\bibitem[{Masciadri {et~al.}(2013)Masciadri, Lascaux, \& Fini}]{Masciadri2013}
Masciadri, E., Lascaux, F., \& Fini, L. 2013, Monthly Notices of the Royal
  Astronomical Society, 436, 1968

\bibitem[{{Masciadri} {et~al.}(2014){Masciadri}, {Lombardi}, \&
  {Lascaux}}]{Masciadri2014}
{Masciadri}, E., {Lombardi}, G., \& {Lascaux}, F. 2014, \mnras, 438, 983

\bibitem[{{Masciadri} {et~al.}(2023){Masciadri}, {Turchi}, \&
  {Fini}}]{Masciadri2023}
{Masciadri}, E., {Turchi}, A., \& {Fini}, L. 2023, \mnras, 523, 3487

\bibitem[{{Munro} {et~al.}(2023){Munro}, {Hansen}, {Travouillon}, {Grosse}, \&
  {Tokovinin}}]{Munro2023}
{Munro}, J., {Hansen}, J., {Travouillon}, T., {Grosse}, D., \& {Tokovinin}, A.
  2023, Journal of Astronomical Telescopes, Instruments, and Systems, 9, 017004

\bibitem[{{N. Morujão}(2023)}]{turlib}
{N. Morujão}. 2023, Turlib: Turbulence estimation library

\bibitem[{{Noll}(1976)}]{Noll1976}
{Noll}, R.~J. 1976, Journal of the Optical Society of America (1917-1983), 66,
  207

\bibitem[{Osborn {et~al.}(2021)Osborn, Townson, Farley, Reeves, \&
  Calvo}]{Osborn2021}
Osborn, J., Townson, M.~J., Farley, O. J.~D., Reeves, A., \& Calvo, R.~M. 2021,
  OPTICS EXPRESS, 29, 6113

\bibitem[{{Osborn} {et~al.}(2018){Osborn}, {Wilson}, {Sarazin}, {Butterley},
  {Chac{\'o}n}, {Derie}, {Farley}, {Haubois}, {Laidlaw}, {LeLouarn},
  {Masciadri}, {Milli}, {Navarrete}, \& {Townson}}]{Osborn2018}
{Osborn}, J., {Wilson}, R.~W., {Sarazin}, M., {et~al.} 2018, \mnras, 478, 825

\bibitem[{{Pepe} {et~al.}(2021){Pepe}, {Cristiani}, {Rebolo}, {Santos},
  {Dekker}, {Cabral}, {Di Marcantonio}, {Figueira}, {Lo Curto}, {Lovis},
  {Mayor}, {M{\'e}gevand}, {Molaro}, {Riva}, {Zapatero Osorio}, {Amate},
  {Manescau}, {Pasquini}, {Zerbi}, {Adibekyan}, {Abreu}, {Affolter}, {Alibert},
  {Aliverti}, {Allart}, {Allende Prieto}, {{\'A}lvarez}, {Alves}, {Avila},
  {Baldini}, {Bandy}, {Barros}, {Benz}, {Bianco}, {Borsa}, {Bourrier},
  {Bouchy}, {Broeg}, {Calderone}, {Cirami}, {Coelho}, {Conconi}, {Coretti},
  {Cumani}, {Cupani}, {D'Odorico}, {Damasso}, {Deiries}, {Delabre},
  {Demangeon}, {Dumusque}, {Ehrenreich}, {Faria}, {Fragoso}, {Genolet},
  {Genoni}, {G{\'e}nova Santos}, {Gonz{\'a}lez Hern{\'a}ndez}, {Hughes},
  {Iwert}, {Kerber}, {Knudstrup}, {Landoni}, {Lavie}, {Lillo-Box}, {Lizon},
  {Maire}, {Martins}, {Mehner}, {Micela}, {Modigliani}, {Monteiro}, {Monteiro},
  {Moschetti}, {Murphy}, {Nunes}, {Oggioni}, {Oliveira}, {Oshagh}, {Pall{\'e}},
  {Pariani}, {Poretti}, {Rasilla}, {Rebord{\~a}o}, {Redaelli}, {Santana
  Tschudi}, {Santin}, {Santos}, {S{\'e}gransan}, {Schmidt}, {Segovia},
  {Sosnowska}, {Sozzetti}, {Sousa}, {Span{\`o}}, {Su{\'a}rez Mascare{\~n}o},
  {Tabernero}, {Tenegi}, {Udry}, \& {Zanutta}}]{ESPRESSO2021}
{Pepe}, F., {Cristiani}, S., {Rebolo}, R., {et~al.} 2021, \aap, 645, A96

\bibitem[{{Perera} {et~al.}(2023){Perera}, {Wilson}, {Butterley}, {Osborn},
  {Farley}, \& {Laidlaw}}]{Perera2023}
{Perera}, S., {Wilson}, R.~W., {Butterley}, T., {et~al.} 2023, \mnras, 520,
  5475

\bibitem[{Roggemann {et~al.}(1996)Roggemann, Welsh, \&
  Hunt}]{roggemann1996imaging}
Roggemann, M.~C., Welsh, B.~M., \& Hunt, B.~R. 1996, Imaging through turbulence
  (CRC press)

\bibitem[{{Salmon} {et~al.}(2009){Salmon}, {Cuillandre}, {Barrick}, {Thomas},
  {Ho}, {Matsushige}, {Benedict}, \& {Racine}}]{Salmon2009}
{Salmon}, D., {Cuillandre}, J.-C., {Barrick}, G., {et~al.} 2009, \pasp, 121,
  905

\bibitem[{Sarazin {et~al.}(2008)Sarazin, Melnick, Navarrete, \&
  Lombardi}]{sarazin2008seeing}
Sarazin, M., Melnick, J., Navarrete, J., \& Lombardi, G. 2008, ESO messenger,
  132, 11

\bibitem[{Southern {et~al.}(2015)Southern, Headquarters,
  Karl-Schwarzschild-Straße, Dorigo, Eso, Navarrete, Sarazin, Vera, Eso, \&
  Vuong}]{ESO_PAPER}
Southern, E., Headquarters, O., Karl-Schwarzschild-Straße, G., {et~al.} 2015,
  European Organisation for Astronomical Research in the Southern Hemisphere
  Programme: PIP Astronomical Site Monitor Data User Manual Change Record from
  previous Version Astronomical Site Monitor Data User Manual

\bibitem[{Southwell(1980)}]{Southwell1980}
Southwell, W. 1980, J. Opt. Soc. Am., 70, 998

\bibitem[{Takato \& Yamaguchi(1995)}]{takato1995spatial}
Takato, N. \& Yamaguchi, I. 1995, JOSA A, 12, 958

\bibitem[{{Tillayev} {et~al.}(2021){Tillayev}, {Azimov}, \&
  {Hafizov}}]{Tillayev2021}
{Tillayev}, Y., {Azimov}, A., \& {Hafizov}, A. 2021, Galaxies, 9, 38

\bibitem[{{Tokovinin}(2021)}]{Tokovinin2021}
{Tokovinin}, A. 2021, \mnras, 502, 794

\bibitem[{Townson {et~al.}(2019)Townson, Farley, de~Xivry, Osborn, \&
  Reeves}]{townson2019aotools}
Townson, M., Farley, O., de~Xivry, G.~O., Osborn, J., \& Reeves, A. 2019,
  Optics express, 27, 31316

\bibitem[{{van Kooten} \& {Izett}(2022)}]{Kooten2022}
{van Kooten}, M. A.~M. \& {Izett}, J.~G. 2022, \pasp, 134, 095001

\bibitem[{{V{\'a}zquez Rami{\'o}} {et~al.}(2012){V{\'a}zquez Rami{\'o}},
  {Vernin}, {Mu{\~n}oz-Tu{\~n}{\'o}n}, {Sarazin}, {Varela}, {Trinquet},
  {Delgado}, {Fuensalida}, {Reyes}, {Benhida}, {Benkhaldoun}, {Garc{\'\i}a
  Lambas}, {Hach}, {Lazrek}, {Lombardi}, {Navarrete}, {Recabarren}, {Renzi},
  {Sabil}, \& {Vrech}}]{Rami2012}
{V{\'a}zquez Rami{\'o}}, H., {Vernin}, J., {Mu{\~n}oz-Tu{\~n}{\'o}n}, C.,
  {et~al.} 2012, \pasp, 124, 868

\bibitem[{{Voitsekhovich}(1995)}]{Voitsekhovich1995}
{Voitsekhovich}, V.~V. 1995, Journal of the Optical Society of America A, 12,
  1346

\bibitem[{Wagner {et~al.}(2022)Wagner, Saxenhuber, Ramlau, \&
  Hubmer}]{WAGNER2022}
Wagner, R., Saxenhuber, D., Ramlau, R., \& Hubmer, S. 2022, Astronomy and
  Computing, 40, 100590

\bibitem[{{Woillez} {et~al.}(2019){Woillez}, {Abad}, {Abuter}, {Aller
  Carpentier}, {Alonso}, {Andolfato}, {Barriga}, {Berger}, {Beuzit}, {Bonnet},
  {Bourdarot}, {Bourget}, {Brast}, {Caniguante}, {Cottalorda}, {Darr{\'e}},
  {Delabre}, {Delboulb{\'e}}, {Delplancke-Str{\"o}bele}, {Dembet}, {Donaldson},
  {Dorn}, {Dupeyron}, {Dupuy}, {Egner}, {Eisenhauer}, {Fischer}, {Frank},
  {Fuenteseca}, {Gitton}, {Gont{\'e}}, {Guerlet}, {Guieu}, {Gutierrez},
  {Haguenauer}, {Haimerl}, {Haubois}, {Heritier}, {Huber}, {Hubin}, {Jolley},
  {Jocou}, {Kirchbauer}, {Kolb}, {Kosmalski}, {Krempl}, {Le Bouquin}, {Le
  Louarn}, {Lilley}, {Lopez}, {Magnard}, {Mclay}, {Meilland}, {Meister},
  {Merand}, {Moulin}, {Pasquini}, {Paufique}, {Percheron}, {Pettazzi}, {Pfuhl},
  {Phan}, {Pirani}, {Quentin}, {Rakich}, {Ridings}, {Riedel}, {Reyes},
  {Rochat}, {Santos Tom{\'a}s}, {Schmid}, {Schuhler}, {Shchekaturov}, {Seidel},
  {Soenke}, {Stadler}, {Stephan}, {Su{\'a}rez}, {Todorovic}, {Valdes},
  {Verinaud}, {Zins}, \& {Z{\'u}{\~n}iga-Fern{\'a}ndez}}]{Woillez2019}
{Woillez}, J., {Abad}, J.~A., {Abuter}, R., {et~al.} 2019, \aap, 629, A41

\bibitem[{Wyngaard(2010)}]{wyngaard2010turbulence}
Wyngaard, J.~C. 2010, Turbulence in the Atmosphere (Cambridge University Press)

\bibitem[{{Zhu} {et~al.}(2023){Zhu}, {Zhang}, {Sun}, {Li}, {Yang}, {He},
  {Weng}, \& {Deng}}]{Zhu2023}
{Zhu}, L., {Zhang}, H., {Sun}, G., {et~al.} 2023, \mnras

\bibitem[{{Ziad} {et~al.}(2004){Ziad}, {Sch{\"o}ck}, {Chanan}, {Troy},
  {Dekany}, {Lane}, {Borgnino}, \& {Martin}}]{Ziad2004}
{Ziad}, A., {Sch{\"o}ck}, M., {Chanan}, G.~A., {et~al.} 2004, \ao, 43, 2316

\end{thebibliography}

\begin{appendix}

\section{   Logarithmic scale of the employed chi-squared weighting factors}
\label{annex: Reconstructed variances}

\begin{figure}[ht]
    \resizebox{\hsize}{!}{\includegraphics{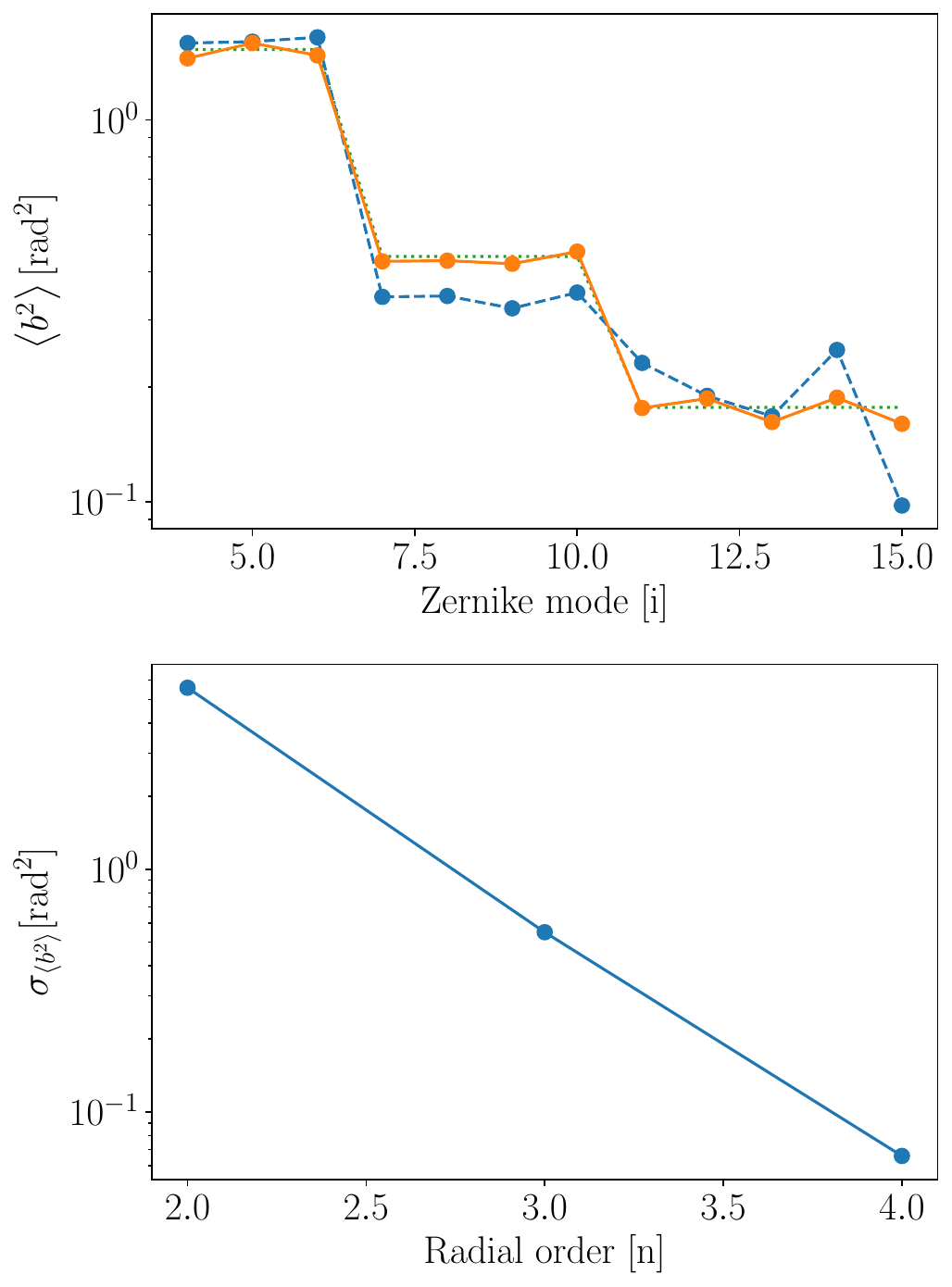}}
    \caption{In the top graph: (i) the dashed line -- the reconstructed variances, $\var{b}{}$, for a simulated telemetry sample; (ii) the dotted line -- the theoretical von Kármán variances of the phase screen turbulence parameters; and (iii) the solid line -- the corrected variances. In the bottom graph: the weight factors, $\sigma_{\varnv{b}{},i}$, obtained from the variance of each radial order of the reconstructed variances presented in the preceding graph.}
    \label{fig: reconstructed coefficients}
\end{figure}

In this appendix, we address the non-constancy of the weighting factors in the chi-squared fit. Through the top plot of Figure \ref{fig: reconstructed coefficients}, we exemplify the fitting of the algorithm to the theoretical variances. The theoretical and corrected variances overlap. We defined the corrected variances, $\var{b}{}_\mathrm{corr}$, as the subtraction of the estimated error from the reconstructed variances, such that
    
    \begin{equation}
        \var{b}{}_\mathrm{corr}  = \var{b}{} - \vec{R} - \vec{\sigma^2_\mathrm{n}}.
    \end{equation}
    
\noindent The approximate logarithmic scale of the squared uncertainty of our data points is illustrated through the bottom plot of the figure, empirically validating the choice of $\sigma_{\varnv{b}{},i}$ as a weight factor to the chi-squared model. 
    
\section{Measurement error estimation from temporal autocorrelation of reconstructed coefficients}
\label{annex: Fusco noise}
Since we assume that $\var{e}{}$ is temporally non-correlated, we can express the reconstructed modal variances through equation (\ref{eq: modal variances}). As such, the total reconstructed variance corresponds to the non-correlated measurement error contribution and remaining terms that depend on atmospheric variances. The temporal autocorrelation of the reconstructed Zernike coefficients, $C_i(\tau)$, for an individual Noll order, $i$, is then given as

\begin{equation}
    C_i(\tau)  = \langle {b_i(t)}{b_i(t+\tau)}\rangle_t = C_{\mathrm{turb},i}(\tau) + \sigma_{\mathrm{n},i}^2\,\delta\tau,
\end{equation}
 
\noindent where $C_{\mathrm{turb},i}(\tau)$ is the temporal autocorrelation of the atmospheric contributions and $ \sigma_{\mathrm{n},i}^2$ is the measurement error. The measurement error is retrieved from $C_i(0)$ through the difference between $C_i(0)$ and $C_{\mathrm{turb},i}(0)$. 

Since the measurement error is only present at $\tau = 0$, we can estimate $C_{\mathrm{turb},i}(0)$ through the polynomial fit of the first points of autocorrelation (excluding $\tau=0$).  We note that $C_{\mathrm{turb},i}(0)$ is estimated through the intercept at $\tau=0$.

\begin{figure}[ht]
    \resizebox{\hsize}{!}{\includegraphics{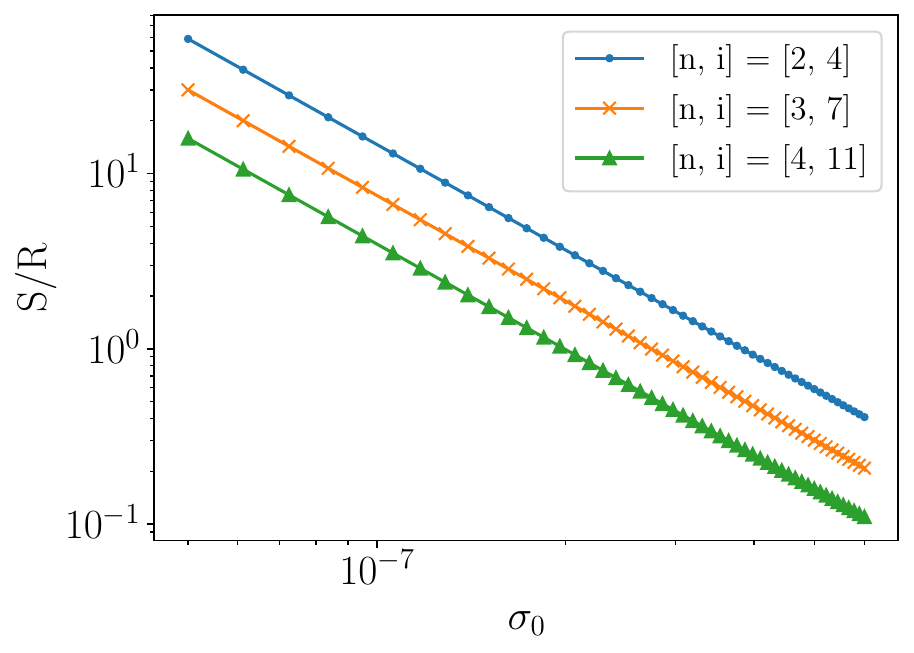}}
    \caption{S/N for individual, $i$, Zernike modes in different radial orders, $n$. The measurement error was verified to be distinct between radial orders.}
    \label{fig: SNR modes}
\end{figure}

The measurement error is introduced as white noise in the slopes, following a known standard deviation, $\sigma$ (c.f. equation (\ref{eq: noise def})). Since the measurement error is obtained from the reconstructed coefficients, we need to take the effects of the reconstruction matrix, $\mat{H}$, into account in the measurement error vector, ${\vec{\sigma^2_{\mathrm{n}}}}$, through equation (\ref{eq: measurement error}). As such, this error cannot be considered constant across Zernike modes. The S/N of individual Zernike modes is defined as

\begin{equation}
    S/N_i = \frac{\varnv{b}{i}}{ \sigma^2_{\mathrm{n},i}},
\end{equation}

\noindent and is represented in Figure \ref{fig: SNR modes} for modes of distinct radial orders, $n$. Due to the non-constant measurement error in the Zernike modes, we estimate the latter as a vector across all reconstructed modes.

\section{Fried parameter estimation performance comparison for different measurement error estimation techniques} 
\label{annex: Fusco noise comparison}

To illustrate the choice of our estimation method, we tested both measurement error estimations:

\begin{itemize}
    \item Simultaneously with the fitting of equation (\ref{eq: full algorithm}), following the specifications outlined in \cite{Andrade2019};
    \item Independently, using the methodology introduced by \cite{Fusco2004}.
\end{itemize}

\begin{figure}[ht]
    \resizebox{\hsize}{!}{\includegraphics{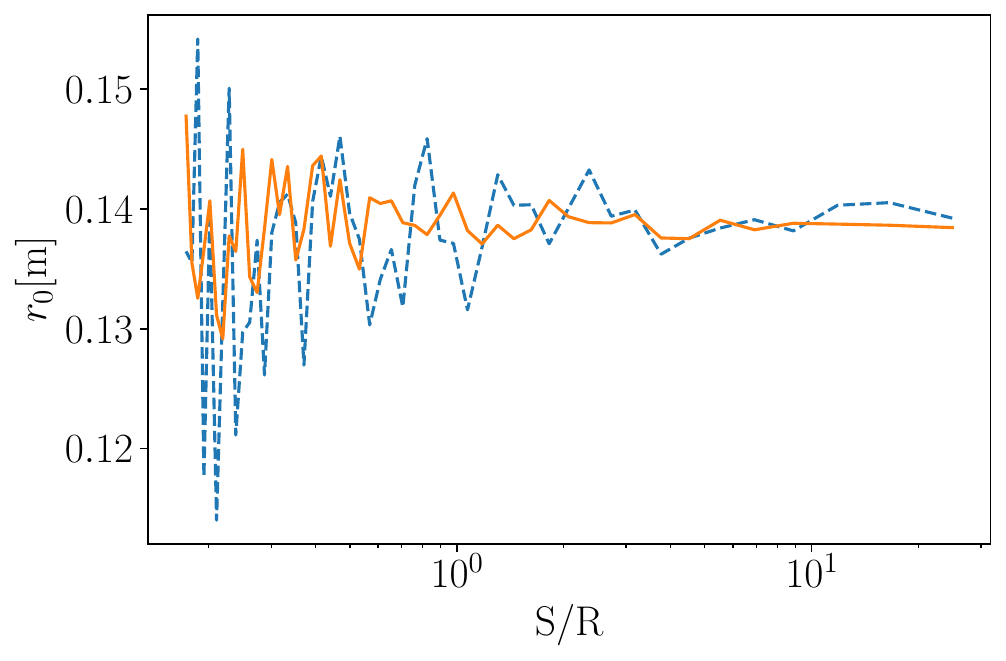}}
    \caption{Estimation of the Fried parameter as a function of S/N, as defined by equation (\ref{eq: SNR}): (i) bold -- autocorrelation of the reconstructed coefficients; and (ii) dashed -- the global fit approach.} 
    \label{fig: SNR relation}
\end{figure}

To illustrate their effect on the convergence of the algorithm, we applied both methods to simulated telemetry samples at different S/N regimes, following the same atmospheric conditions expressed in Section \ref{subsec: simulation results}. The results are illustrated in Figure \ref{fig: SNR relation}. We find that the Fried parameter estimation by the autocorrelation method is more stable in the low S/N regime. As such, the estimation of the measurement error in the low S/N regime is improved, while performing equally as well in the high S/N regime. Additionally, by removing the measurement error from the chi-squared minimisation, we reduced the number of fitting variables from $\vec{p} = (r_0, \mathcal{L}_0, \sigma_0$) to $\vec{p} = (r_0, \mathcal{L}_0)$.

\end{appendix}

\end{document}